# On Small Beams with Large Topological Charge

Mario Krenn[1,2,*], Nora Tischler[1,2,3,*], Anton Zeilinger[1,2]

[1]Vienna Center for Quantum Science and Technology (VCQ), Faculty of Physics, University of Vienna, Boltzmanngasse 5, A-1090 Vienna, Austria.
[2]Institute for Quantum Optics and Quantum Information (IQOQI), Austrian Academy of Sciences, Boltzmanngasse 3, A-1090 Vienna, Austria.
[3]Centre for Engineered Quantum Systems, Department of Physics & Astronomy, Macquarie University, NSW 2109, Sydney, Australia.
[*]these authors contributed equally to this work.

**Light beams can carry a discrete, in principle unbounded amount of angular momentum. Examples of such beams, the Laguerre-Gauss modes, are frequently expressed as solutions of the paraxial wave equation. There, they are eigenstates of the orbital angular momentum (OAM) operator. The paraxial solutions predict that beams with large OAM could be used to resolve arbitrarily small distances – a dubious situation. Here we show how to solve that situation by calculating the properties of beams free from the paraxial approximation. We find the surprising result that indeed one can resolve smaller distances with larger OAM, although with decreased visibility. If the visibility is kept constant (for instance at the Rayleigh criterion, the limit where two points are reasonably distinguishable), larger OAM does not provide an advantage. The drop in visibility is due to a field in the direction of propagation, which is neglected within the paraxial limit.**

Laguerre-Gauss (LG) modes can be specified by two mode numbers, $\ell$ and $n$, which are the orbital angular momentum (OAM) mode number or topological charge, and the radial mode number, respectively [1]. For the rest of this note, we consider beams with the radial mode number $n=0$ and look at the transverse beam patterns in the plane where z=0.

Laguerre-Gauss beams can be found as solutions of the paraxial wave equation and described in cylindrical coordinates $(r, \varphi)$ by

$$LG_\ell(r,\varphi) = N \cdot \left(\frac{r}{w_0}\right)^{|\ell|} \cdot e^{-\frac{r^2}{w_0^2}} \cdot e^{-i\ell\varphi}, \qquad (1)$$

where $w_0$ is the beam waist and N is a normalization constant. The intensity is given by

$$I_\ell(r,\varphi) = |LG(r,\varphi)|^2, \qquad (2)$$

which results in an intensity ring for $\ell > 0$. The intensity maximum of the ring in the radial direction is at

$$r_{max} = \sqrt{\frac{\ell}{2}} w_0. \qquad (3)$$

The radius of maximum intensity scales with the square-root of the OAM [2, 3]. Superpositions of two LG-modes with opposite OAM have the same radial dependence (consequently, the intensity maximum is at the same $r_{max}$), but they exhibit intensity oscillations in the azimuthal direction with $2\ell$ intensity maxima and minima. The distance between two petals (intensity maxima) is therefore

$$\Delta = \frac{\pi w_0}{\sqrt{2\ell}}. \qquad (4)$$

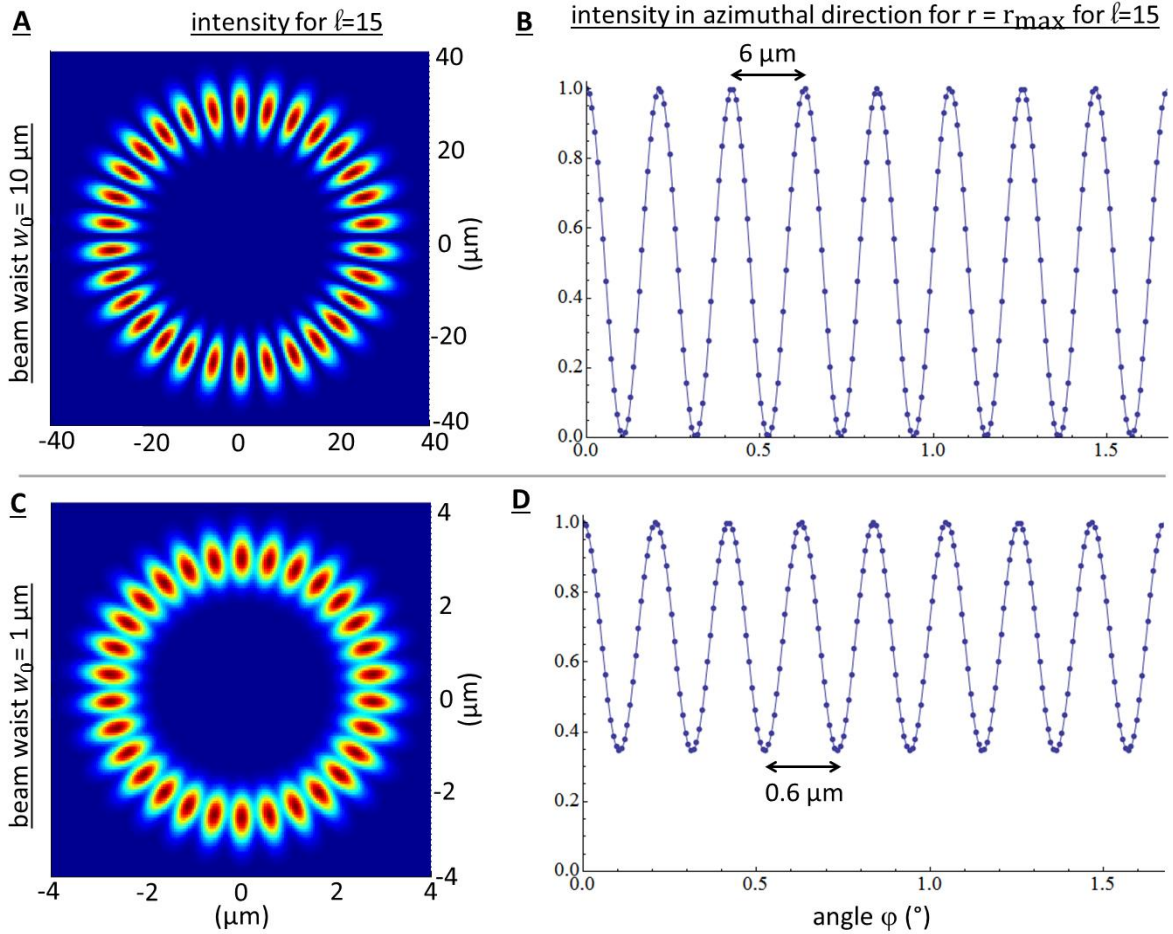

**Figure 1**: LG beams with λ=800nm and $\ell$ =15 with different beam waist $w_0$. <u>A</u>: intensity profile of the LG mode with $w_0$=10 μm. <u>B</u>: Intensity in azimuthal direction for r=$r_{max}$. The visibility of the fringes is close to unity, which is very close to the paraxial case. <u>C</u> & <u>D</u>: The same properties for a LG beam with $w_0$=1 μm. The minima are washed out significantly, the visibility drops to roughly 49%, which is a pure non-paraxial effect.

For a fixed beam waist, the distance between two petals becomes smaller with increasing $\ell$. A potential application of these superposition beams is the use of the azimuthally varying intensity pattern to probe small structures. Then the distance Δ between two maxima will have a bearing on the achievable resolution. Based on the paraxial solution from eqn. (1) it seems as if we should be able to resolve arbitrarily small structures since Δ can be decreased by increasing $\ell$. We could resolve in the sub-wavelength regime, but also further in the sub-atomic or in the extreme case even Planck-length regime (which would be found for $\ell \sim 10^{62}$, where coincidentally the mode has the diameter of the observable universe). This would obviously constitute a very curious situation and the question is how it can be resolved.

The calculation above is based on the paraxial approximation, which is only valid for sufficiently large beams: It is known that the paraxial wave equation is a zero-order approximation of the Maxwell's equations with terms of the order $\alpha = \mathcal{O}\left(\frac{\lambda}{w_0}\right)$ ignored, where λ is the wavelength [4-6]. Here, we test if these considerations withstand a more rigorous analysis free from the paraxial approximation. We use forward-propagating LG modes which are full solutions of Maxwell's equations. In order to calculate full solutions of LG beams, we use two distinct methods proposed in the literature. The first method is based on the elegant framework of the Riemann-Silberstein vector [7, 8]. The finite energy, analytical solution of LG beams is derived in [9]. The second method to calculate full Maxwell solutions of LG beams is based on the aplanatic lens model [10-12] which is a standard way to describe focused fields. The familiar paraxial properties of the modes are recovered by the solutions of both methods. For small beams with large OAM the two methods show the same behavior, which deviates from the predictions obtained using the paraxial approximation. Specifically, both methods unanimously show that the visibility of the intensity fringes in the azimuthal direction decreases. This is due to an increasing field component in the direction of propagation which is out of phase with the transverse components. This solves the curious situation explained above. Usually, the smallest distance one can resolve is given by the diffraction limit $d = \frac{\lambda}{2}$. Although the LG solutions in the Riemann-Silberstein formalism are analytic, they are more difficult to interpret with respect to the diffraction limit as they are not monochromatic. Therefore, we continue with the aplanatic lens model.

The aplanatic lens model can be used to calculate strongly focused fields that are obtained by the use of a microscope objective or similar. Based on properties of the focusing optics and the incident field, the model provides the focused electric field, which is a solution of the full Maxwell's equations. The properties specifying the focusing optics are the focal length, the numerical aperture (NA), and the transmission coefficients for s- and p-polarization. We assume an ideal microscope objective by setting the NA to 1 and by letting the s and p transmission coefficients be unity, which is the goal of antireflection coatings. For the incident field, we use the LG modes specified by (1), with circular polarization $\hat{u}(r,\sigma) = \frac{1}{\sqrt{2}}(\hat{x} + i\sigma\hat{y})$. Such a circularly polarized collimated beam is to a very good approximation a helicity eigenstate (with eigenvalue $\sigma$), and the helicity (i.e. circular polarization of each plane wave composing the total field) is preserved throughout the focusing process owing to the equal s- and p-transmission coefficients [12].

The aplanatic lens model essentially associates real space field coefficients of the input beam with plane wave decomposition amplitudes of the focused field, and gives us the following output field [13]:

$$\boldsymbol{E}_{\ell,\sigma}(\boldsymbol{r},t) = \int_0^{k_{max}} g(k_r) \cdot k_r \, dk_r \int_0^{2\pi} dk_\phi \, LG(k_r, k_\phi, w_0) \cdot e^{i(\boldsymbol{k}\cdot\boldsymbol{r}-\omega t)} \cdot \hat{\boldsymbol{u}}(\boldsymbol{k},\sigma)$$

$$LG(k_r, k_\phi, w_0) = \sqrt{\frac{w_0^2}{2\pi |\ell|!}} \, e^{i\ell k_\phi} \cdot \left(\frac{w_0 k_r}{\sqrt{2}}\right)^{|\ell|} \cdot e^{-\frac{w_0^2 \cdot k_r^2}{4}} \cdot e^{-i\frac{|\ell|\pi}{2}}$$

$$\hat{\boldsymbol{u}}(\boldsymbol{k},\sigma) = \frac{e^{i\sigma k_\phi}}{\sqrt{2}k\sqrt{k_x^2 + k_y^2}} \begin{pmatrix} -k_x k_z + \sigma i k k_y \\ -k_y k_z - \sigma i k k_x \\ k_x^2 + k_y^2 \end{pmatrix}$$

(5)

Here, $w_0 = \frac{f \cdot \lambda}{w_{in} \cdot \pi}$ is the beam waist after focusing the incoming beam (with waist $w_{in}$) with a lens of focal length f. $(k_r, k_\phi, k_z)$ and $(k_x, k_y, k_z)$ are cylindrical and Cartesian coordinates in momentum space, with $k = \sqrt{k_r^2 + k_z^2} = \sqrt{k_x^2 + k_y^2 + k_z^2} = \frac{2\pi}{\lambda}$ denoting the wave number and $\lambda$ being the optical wavelength. The integration of $k_r$ is cut off at $k_{max} = k$ which implies a numerical aperture of the focusing objective of 1, and effectively avoids evanescent waves as we are only interested in propagating fields. The factor $g(k_r) = \sqrt[4]{1 - (k_r/k)^2}$ comes from energy flux conservation during focusing, and is responsible for a damping of high radial k-components at very strong focusing.

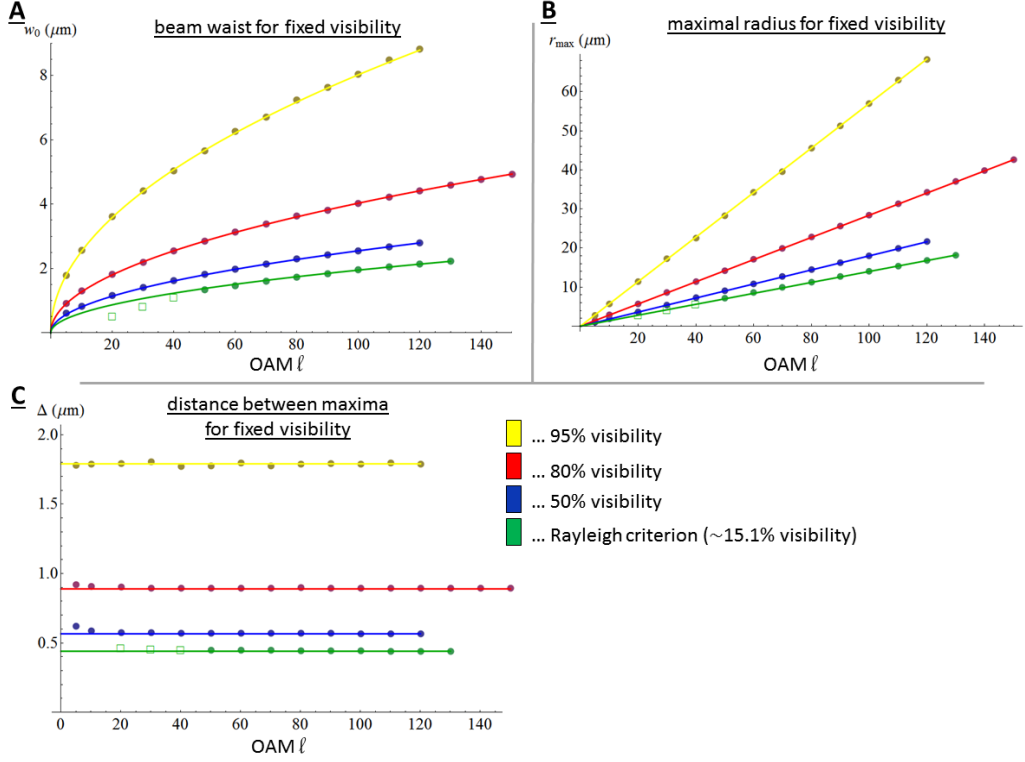

**Figure 2**: A: If we fix the visibility of the intensity in the azimuthal direction, we find that the beam waist $w_0$ increases as a square root of the OAM. B: This leads to a linear scaling of the maximum radius of the intensity pattern. C: As a consequence, the distance Δ stays constant. For all plots the points represent calculated values and the line stands for square-root or linear interpolation. The cases where significant portions of the beam are cut off due to the cut-off $k_{max}$ from eq.(5) are indicated by squares.

The helicity is given by $\sigma$, which can be +1 or -1 for left- or right-circular polarization. $\hat{u}(k,\sigma)$ is a normalized circular-polarization vector [8] in Cartesian coordinates. As we would like to produce superposition of LG modes, we can simply add two fundamental solutions: $E_{\pm\ell,\sigma}(r,t) = \frac{1}{\sqrt{2}}\left(E_{+\ell,\sigma}(r,t) + E_{-\ell,\sigma}(r,t)\right)$.

Now we study the intensity distribution of very small beams with non-zero OAM superposition. The intensity can simply be calculated as $I_\ell(r,\varphi) = \sum_{i=\{x,y,z\}} Re\{E_i\}^2 + c^2 Re\{B_i\}^2$, where $Re\{\}$ denotes the real part. The intensity $I_\ell(r,\varphi)$ is related to the full energy of the electromagnetic field. It is also meaningful in quantum physics due to its interpretation as the probability density for the case of single photons [8][14].

We fix our wavelength to $\lambda = 800$ nm. In Fig.1 we plot the intensity of a beam with $\ell$ =15, which leads to 30 petals in the ring. In 1A-B, the beam waist is $w_0$=10 µm, whereas in 1C-D $w_0$=1 µm is used. In the smaller beam, the intensity minima are significantly filled in due to a large field component in z-direction (in the example, the maximum of the field in z-direction is roughly 32% of the full intensity's maximum). The z-component is shifted azimuthally by exactly half a period compared to the x- and y-components. The visibility (defined as $vis = \frac{I_{max}-I_{min}}{I_{max}+I_{min}}$ at the radial position of the intensity maximum) drops from almost unity to 48.6%.

In Fig. 2, we show the OAM dependence of the beam waist $w_0$, the maximum radius $r_{max}$ and the distance between different maxima Δ. If no other restrictions are applied and $w_0$ remains constant, indeed $r_{max}$ scales like the square root of the OAM, therefore Δ decreases – exactly as predicted by the paraxial solution. However, if one takes into account the decrease of the visibility, which is a significant measure for resolution, the situation changes. When the visibility is fixed (by adjusting $w_0$), the maximum radius $r_{max}$ increases linearly with the OAM, which leads to a constant behaviour of the Δ. The behavior is analyzed for different visibilities of superposition fringes: 95%, 80% 50% and 15.1%. The last one resembles the Rayleigh criterion, the limit at which two points can be reasonably distinguished. These results show that OAM superpositions cannot be used to decrease the distance between two maxima while keeping the minima small – which might be a significant result for OAM-based resolution techniques.

These conclusions are based on the natural choice of using the radius at which the total energy of the field is maximal. One interesting matter is the amplitude of the fringes. In all cases presented above we have analyzed the visibility in the radial position of maximal intensity. We also analyzed azimuthal visibilities for r≠$r_{max}$, and see that the visibility decreases for smaller r (in regions where there is still a considerable amount of intensity) [13]. However, it is known that situations can exist where high frequency oscillations with perfect visibilities can be achieved, in places where the intensity is exponentially small [15]. We conjecture that there is a criterion which jointly takes into account the overall amplitude of the fringes and their visibility, and indicates the usefulness for resolution. A different workaround for the limitations explained in this work could be the use of a method that is only sensitive to certain parts of the electromagnetic field, such as the electric energy or the transverse components of the field.

Our result shows that eq.(1) leads to incorrect predictions in the regime of small beams with large $\ell$. This is for two distinct reasons. First of all, the field component in the z-direction, which is neglected in the paraxial approximation, causes reduced visibilities, with possible implications for imaging applications or optical lattices [16]. Secondly, limiting the field to propagating modes imposes a cut-off in the radial momentum components, which becomes significant for even smaller beam waists or larger $\ell$ (see eq. (5)). A fascinating question is the behaviour of matter waves with large orbital angular momentum. As they are described by the Schrödinger equation, which has the same form as the paraxial wave equation, the visibility issue in OAM superpositions does not apply – indicating an interesting difference between propagation of photons and matter waves [17, 18]. However, the physical constraint on the maximal transverse momentum for propagating modes is valid for matter waves and poses a limitation on their use for resolution applications, similarly as in the case of photons. The paraxial wave equation, for which eq.(1) is a solution, is an approximation of an optical Dirac equation [19], in a formally very similar way as the Schrödinger equation is an approximation of the massive Dirac equation. It would be interesting to investigate whether similar non-paraxial effects presented here exist in some form for relativistic matter waves as well.

## Acknowledgement

The authors thank Robert Boyd, Peter Banzer, Ebrahim Karimi, Miles Padgett, Gabriel Molina-Terriza and Iwo Bialynicki-Birula for interesting discussions. This project was supported by Austrian Academy of Sciences (ÖAW), the European Research Council (SIQS Grant No. 600645 EU-FP7-ICT), and the Austrian Science Fund (FWF) with SFB F40 (FOQUS). NT was supported by the Australian Research Council's Centre of Excellence for Engineered Quantum Systems (EQuS).

# Supplementary Information

## 1) Derivation of the focused LG field

To arrive at (5), we use the LG modes of Eq. (1) with circular polarization $\hat{u}(r,\sigma) = \frac{1}{\sqrt{2}}(\hat{x} + i\sigma\hat{y})$ as $E_{in}(r,t) = E_{in}(r,\varphi,0,t)$. This means that the beam waist of the paraxial beam is at the input of the microscope objective. Then the coefficients of the plane wave decomposition for the focused field are obtained by making the substitutions $r \to \frac{fk_r}{k}$, $\varphi \to k_\phi$ for the coordinates of the input field, and multiplication by the energy flux conservation factor $g(k_r) = \sqrt[4]{1-(k_r/k)^2}$.

The polarization vectors in the plane wave decomposition can be obtained as follows: We start with a helicity eigenstate and our aplanatic lens conserves that helicity, which means that each plane wave of the focused field must have circular polarization with the same handedness as the incident field. $\hat{u}(k,\sigma)$ from Eq. (5) fulfills these requirements: It is a normalized circular-polarization vector [8] in Cartesian coordinates, which is tilted such that it is transverse to the momentum vector $k$, and has its handedness specified by $\sigma$. Specifically, $\hat{u}(k,\sigma)$ can be derived by starting with $\hat{u}(r,\sigma) = \frac{1}{\sqrt{2}}(\hat{x} + i\sigma\hat{y})$, and rotating this vector by $\arcsin(k_r/k)$ about the y axis, and by $k_\phi$ about the z axis. Then an $e^{i\sigma k_\phi}$ factor is applied in order to attain the correct total angular momentum J = ℓ + σ, which is the total angular momentum of the input beam that is conserved by the cylindrically symmetric focusing optics.

The above construction results in the plane wave decomposition

$$E_{\ell,\sigma}(k_r, k_\phi) = g(k_r) \cdot LG(k_r, k_\phi, w_0) \cdot \hat{u}(k,\sigma), \tag{S1}$$

with $g(k_r)$ and $LG(k_r, k_\phi, w_0)$ as specified in the main text. The final step consists in a Fourier transform to obtain the real space field of Eq. (5).

## 2) Visibility vs. Radius

In the main text, we analysed the visibility in azimuthal direction at the radial position where the intensity is maximal. This is a natural choice if we are interested, for instance, in the usefulness for resolution. However, we can also look at different radii and analyse the visibility there. In Figure S1, we calculate the visibilities for different regions of the beam, for different OAM ℓ and different beam waists w₀.

The visibility is calculated for all radii where the intensity is at least 0.1% of the maximum intensity – excluding regions of negligible intensity, which are expected to be unusable in imaging applications. We find that for smaller radii, the visibility only decreases. For larger radii, it increases.

In order to calculate the visibility in other regimes where the intensity is even lower, robust numerical methods need to be used in order to deal with exponentially small intensities.

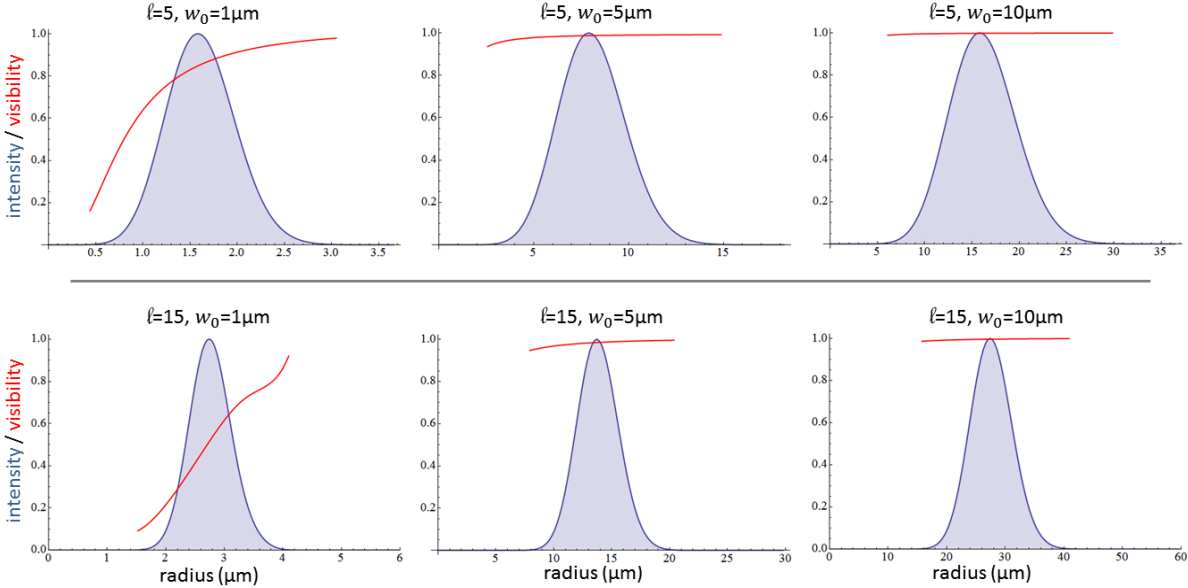

**Figure S1**: The visibility of beams with different waists and different OAM is calculated as a function of the radial coordinate (not only at the radius of maximum intensity – as used in the main text). The blue shape represents the intensity of the LG mode in radial direction. The red line shows the visibility.